# Variation between Antiferromagnetism and Ferrimagnetism in NiPS₃ by Electron Doping

*Mengjuan Mi, Xingwen Zheng, Shilei Wang, Yang Zhou, Lixuan Yu, Han Xiao, Houning Song, Bing Shen, Fangsen Li, Lihui Bai, Yanxue Chen[*], Shanpeng Wang[*], Xiaohui Liu[*], Yilin Wang[*]*

M.J. Mi, L.X. Yu, H. Xiao, Y.L. Wang
School of Microelectronics, Shandong Technology Center of Nanodevices and Integration, State Key Laboratory of Crystal Materials, Shandong University, Jinan 250100, China
E-mail: yilinwang@email.sdu.edu.cn

X.W. Zheng, Y. Zhou, H.N. Song, L.H. Bai, Y.X. Chen, X.H. Liu
School of Physics, Shandong University, Jinan 250100, China
E-mail: cyx@sdu.edu.cn; liuxiaohui@sdu.edu.cn

S.L. Wang, S.P. Wang
State Key Laboratory of Crystal Materials, Institute of Crystal Materials, Shandong University, Jinan, 250100, China
E-mail: wshp@sdu.edu.cn

B. Shen
School of Physics, Sun Yat-Sen University, Guangzhou 510275, China

F.S. Li
Vacuum Interconnected Nanotech Workstation, Suzhou Institute of Nano-Tech and Nano-Bionics, Chinese Academy of Sciences, Suzhou 215123, China

Keywords: two-dimensional magnetic materials, organic cations intercalation, carrier doping, magnetic order transition

How to electrically control magnetic properties of material is promising towards spintronic applications, where the investigation of carrier doping effects on antiferromagnetic (AFM) materials remains challenging due to their zero net magnetization. In this work, we find electron doping dependent variation of magnetic orders of a two-dimensional (2D) AFM insulator NiPS₃, where doping concentration is tuned by intercalating various organic cations into the van der Waals gaps of $NiPS_3$ without introduction of defects and impurity phases. The doped $NiPS_3$ shows an AFM-ferrimagnetic (FIM) transition at a doping level of 0.2-0.5 electrons/cell and a FIM-AFM transition at a doping level of ≥0.6 electrons/cell. We propose that the found phenomenon is due to competition between Stoner exchange dominated inter-chain ferromagnetic order and super-exchange dominated AFM order at different doping level. Our studies provide a viable way to exploit correlation between electronic structures and magnetic properties of 2D magnetic materials for realization of magnetoelectric effect.

## 1. Introduction

Since the first experimental observation of intrinsic ferromagnetism in monolayer $CrI_3$[1] and bilayer $Cr_2Ge_2Te_6$,[2] van der Waals (vdW) magnetic materials have attracted extensive attention in both fundamental research and practical applications.[3-9] The field of two-dimensional (2D) magnetic materials grows rapidly and various types of 2D magnetic materials have been discovered and synthesized, such as ferromagnetic (FM) magnets including $Cr_2Ge_2Te_6$,[2] $CrBr_3$,[10] $Fe_3GeTe_2$,[11-12] $Fe_5GeTe_2$,[13] monolayer $VSe_2$,[14] etc., and antiferromagnetic (AFM) magnets including $CrCl_3$,[15] transition metal phosphorous trichalcogenides $MPX_3$ (M = Mn, Fe, Ni; X = S, Se) ,[16-19] $CrPS_4$,[20] $MnBi_2Te_4$,[21] etc. The distinct spin-dependent properties of these materials provide a promising platform for the discovery and study of new quantum phenomena and design of novel spintronic devices.

Due to the ultrathin thickness and weak interlayer vdW interaction of 2D magnetic materials, their magnetic properties, such as Curie temperature, magnetic anisotropy, saturation magnetization, and coercive force can be effectively modulated by magnetic field,[1] electric field,[22] strain,[23] electrostatic doping,[11, 24-27] and ion intercalation,[28] etc. Several studies have demonstrated the carrier doping dependent changes in magnetic properties of $CrI_3$,[24] $Cr_2Ge_2Te_6$,[26, 28] and $Fe_3GeTe_2$,[11] which are attributed to carrier doping induced change on exchange interaction due to orbital occupation of transition metal atoms in these materials. For example, in monolayer $CrI_3$, the saturation magnetization, coercive force and Curie temperature increase (decrease) linearly for hole (electron) doping.[24] In $Cr_2Ge_2Te_6$, carrier doping significantly increases the Curie temperature ($T_c$) from 61 K to over 200 K and switches the magnetic easy axis from out-of-plane to in-plane at an electron density of $>1 \times 10^{14}/cm^2$.[26, 28] In $Fe_3GeTe_2$, $T_c$ in tri-layer is enhanced from 100 K to over room temperature at an electron density of approximately $10^{14}/cm^2$.[11] Furthermore, electrostatic doping induces the antiferromagnetic-ferromagnetic (AFM-FM) transition in bilayer $CrI_3$ at a critical electron density of $2.5 \times 10^{13}/cm^2$.[24] The modulation of magnetic properties by carrier doping serves as a viable tool for realizing effective magnetoelectric coupling and is promising for designing electric-field controlled spintronic devices. However, among these impressive studies, experimental realization of carrier doping induced magnetic order transition in 2D magnetic materials has rarely been reported,[24] and the mechanism responsible for the magnetic transition is not clear and needs to be further explored.

Among various magnetic materials, $MPX_3$ compounds are of great interest for their rich variety of electronic and magnetic properties depending on the role of the transition metal elements: Heisenberg-type $MnPS_3$, Ising-type $FePS_3$, and XY or XXZ-type $NiPS_3$.[16-19, 29-30] $MPX_3$ compounds have been theoretically predicted to exhibit strong charge-spin coupling[31-32] and carrier doping dependent AFM-FM transition,[33-34] while experimental studies are lack.

In this work, we report on the electron doping dependent AFM-ferrimagnetic (FIM) transition in a self-doped AFM insulator $NiPS_3$, where the electron doping is realized by intercalating organic cations into the van der Waals gaps.[35] Intercalation has been demonstrated to be an effective method to dope electrons into 2D materials and modulate their electronic,[36-37] optical,[38] magnetic,[28] superconductivity,[39-40] thermal conductivity,[41] thermoelectric,[42] and energy storage properties.[43] We found variation of magnetic order from AFM to FIM and then AFM with the increasing of electron concentration. Our experimental results are consistent with theoretical analysis based on the first-principles calculations. We propose that such AFM-FIM transition at moderate electron doping level originates from the Stoner exchange due to the self-doped feature of $NiPS_3$ which provides effective inter-chain itinerant channel, and the FIM-AFM transition at high electron doping level is due to the rise of super-exchange. Our studies provide a new insight into the carrier doping tuned magnetic transition in 2D magnetic materials.

## 2. Results and Discussion

### 2.1 Preparation of organic cations intercalated NiPS₃

**Figure 1**a shows the crystal and magnetic structures of $NiPS_3$.[29-30, 44-45] Bulk $NiPS_3$ has a monoclinic structure (*C2/m*), where two P atoms (P-P pair) are covalently bonded to six S atoms to form a $(P_2S_6)^{4-}$ anion complex, and each Ni atom carries a +2 electronic ionization state and lies on a honeycomb lattice in the *ab* plane. The layers in the *ab* plane are coupled by weak vdW interactions along the *c*-axis, and the interlayer distance is 6.34 Å. Below Néel temperature

($T_N \sim 150$ K), magnetic moments are aligned mostly in the *ab* plane (along the *a*-axis direction) with a small out-of-plane component, and each $Ni^{2+}$ ion is coupled ferromagnetically to two of the nearest neighbors and antiferromagnetically to the third one, forming zigzag ferromagnetic chains (parallel to the *a*-axis) coupled antiferromagnetically to each other along the *b*-axis direction (zigzag AFM order, upper panel in Figure 1a). The non-equivalent $Ni^{2+}$ ions in two adjacent ferromagnetic zigzag chains are marked as Ni(1) and Ni(2), respectively.

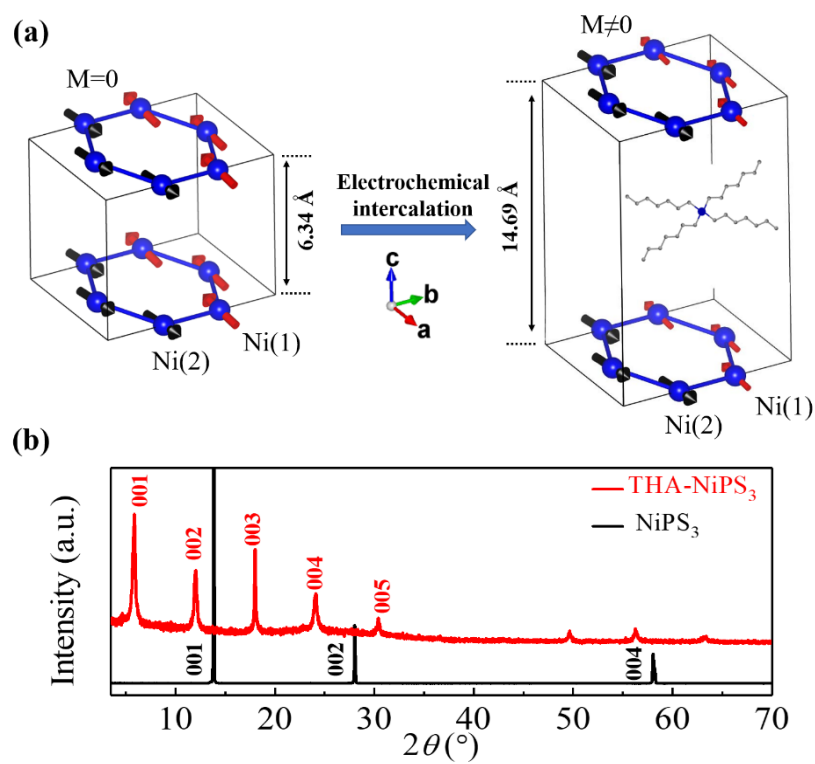

**Figure 1.** Structures of pristine $NiPS_3$ and intercalated THA-$NiPS_3$. a) Structures of $NiPS_3$ (left) and $THA^+$ cations intercalated $NiPS_3$ (THA-$NiPS_3$, right). Arrows indicate the orientation and size of magnetic moments of Ni atoms, and the Ni atoms in two adjacent zigzag ferromagnetic chains are marked as Ni(1) and Ni(2), respectively. b) XRD patterns of pristine $NiPS_3$ and intercalated THA-$NiPS_3$.

Intercalation offers a versatile approach for tuning charge carriers in 2D materials due to the charge transfer between guest intercalants and host 2D materials.[35] Because of the remarkable wide vdW gap in $MPX_3$, guest intercalants can be easily intercalated into the vdW gaps, and these compounds have been extensively studied as cathodic materials in lithium batteries.[46] High-quality $NiPS_3$ bulk crystals were used as the host materials, and electrochemical intercalation method was adopted to insert organic cations, such as $THA^+$, into the vdW gaps of $NiPS_3$, as shown in Figure 1a. The successful intercalation is confirmed by X-ray diffraction (XRD) measurement as shown in Figure 1b, where the obvious shift of diffraction peaks to smaller angles indicates that the interlayer distance is expanded to 14.69 Å.

## 2.2 Determination of ferrimagnetism in intercalated THA-NiPS$_3$

**Figure 2** shows the temperature-dependent magnetization (*M-T*) and field-dependent magnetization (*M-H*) for pristine NiPS$_3$ and intercalated THA-NiPS$_3$ with a magnetic field applied along directions parallel to the *ab* plane and perpendicular to the *ab* plane (*c\**). The overall behavior of pristine NiPS$_3$ is consistent with previous results, and a typical AFM characteristic is observed.[29-30] No substantial difference is observed between the field-cooled (FC) and zero-field cooled (ZFC) measurements. The broad maximum at elevated temperatures is related to the short-range spin correlations. The Néel temperature, $T_N$, defined from the sharp peak in the derivative d*M*/d*T* for *H // ab* is ~150 K (inset of Figure 2a). Below $T_N$, with respect to the magnetization for *H // c\**, the magnetization for *H // ab* decreases rapidly and reaches a smaller value at low temperature (Figure 2a), indicating that the magnetic moments lie mostly in the *ab* plane. A linear dependence of *M vs H* for *H // c\** is observed, while a clear upturn in *M vs H* for *H // ab* above a critical field $\mu_0H$~5 T is observed (Figure 2b). The upturn in the magnetization at high field may be attributed to the spin-flop transition in pristine NiPS$_3$.[47]

After intercalation, an obvious FIM characteristic (discuss in detail later) with a Curie temperature $T_c$~100 K is observed in THA-NiPS$_3$. As shown in Figure 2c,d, for *H // ab*, the magnetization increases rapidly below 100 K, and an obvious magnetic hysteresis loop is observed at a low temperature, which provides an unambiguous identification of FIM order. For *H // c\**, the magnetization is significantly smaller, and a linear dependence of *M vs H* is observed, which suggests that the large magnetic anisotropy still remains in intercalated THA-NiPS$_3$, and the FIM easy axis still lies in the *ab* plane. With increasing temperature (Figure 2e,f), the magnetic hysteresis loop becomes less obvious, and both remnant magnetization and coercive field become smaller monotonically. The *M-H* curve shows a linear dependence at *T* = 100 K (inset in Figure 2e), indicating a FIM-paramagnetic transition, which is consistent with the *M-T* curves in Figure 2c.

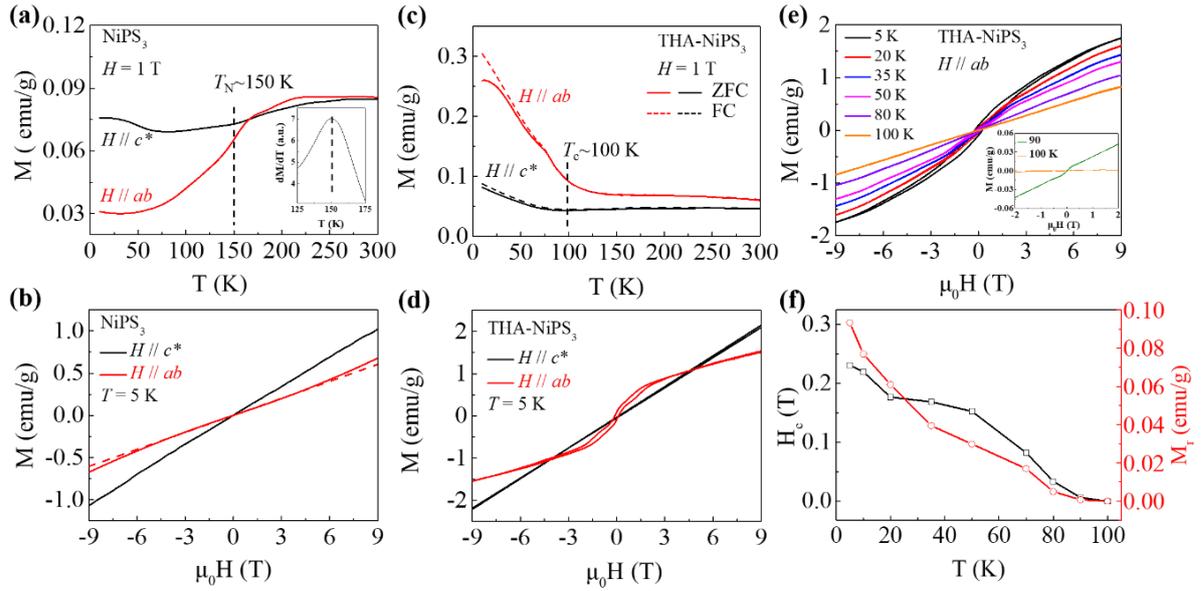

**Figure 2.** Magnetic properties of pristine NiPS₃ and intercalated THA-NiPS₃. a, b) Temperature dependence of magnetization (*M-T*, a) in zero-field cooled (ZFC) mode and field dependence of magnetization (*M-H*, b) of pristine NiPS₃ under magnetic fields *H // ab* (red) and *H // c** (black). c, d) Temperature dependence of magnetization (*M-T*, c) and field dependence of magnetization (*M-H*, d) of intercalated THA-NiPS₃ under magnetic fields *H // ab* (red) and *H // c** (black). The solid and dashed lines in **c** represent zero-field cooled (ZFC) and field cooled (FC) data, respectively. e) Isothermal magnetization of intercalated THA-NiPS₃ under a magnetic field *H // ab* at different temperatures. Inset shows enlarged *M-H* data at 90 K and 100 K after subtracting a linear fitting. f) Extracted remnant magnetization $M_r$ (red) and coercive field $H_c$ (black) of intercalated THA-NiPS₃ as a function of temperature.

## 2.3 Electron doping in intercalated THA-NiPS₃ without introduction of defects and impurity phases

What causes the magnetic transition from AFM in pristine NiPS₃ to FIM in intercalated THA-NiPS₃? Whether such AFM-FIM transition originates from defects or impurity phases? To solve these puzzles, detailed structural and chemical characterizations were carried out. **Figure 3** shows typical morphologies of exfoliated pristine NiPS₃ and intercalated THA-NiPS₃ thin flakes. The intercalated THA-NiPS₃ has a flat and smooth surface, and no obvious defects are observed. After intercalation, the root-mean-square (RMS) roughness slightly increases from 0.2 nm (pristine NiPS₃) to 0.35 nm (THA-NiPS₃). A step height of ~1.6 nm is observed for THA-NiPS₃ (Figure S1, Supporting Information), which is consistent with the XRD results. We provided the first experimental evidence showing that the organic cations intercalated 2D materials host a platform for the atomic-scale investigation by scanning probe techniques.

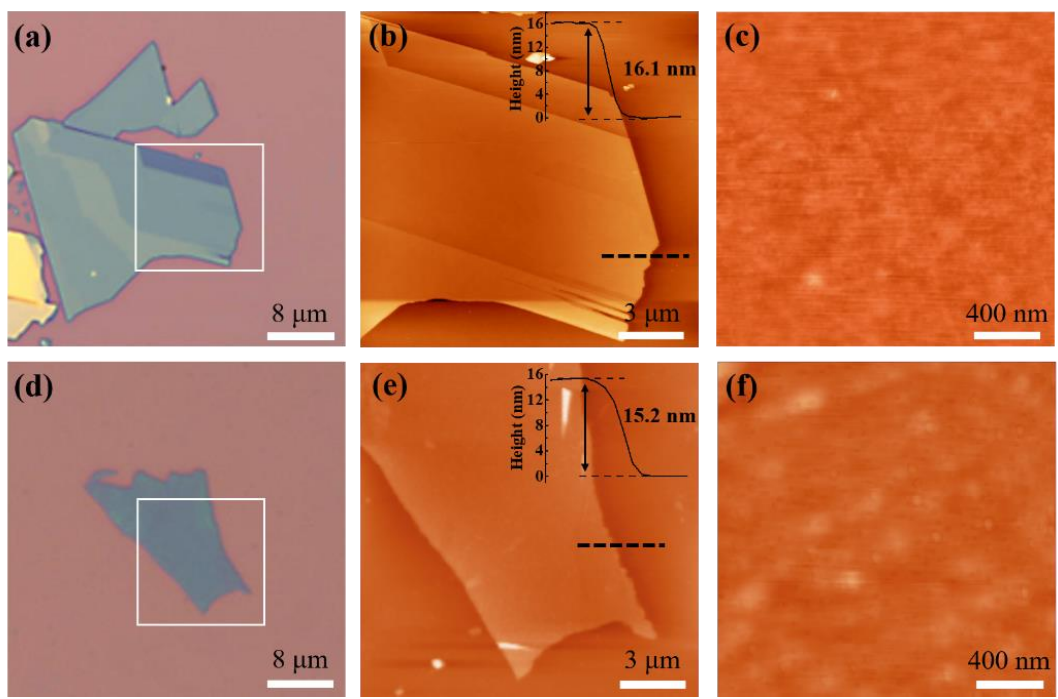

**Figure 3.** Surface morphologies of pristine $NiPS_3$ and intercalated $THA-NiPS_3$. a-c) Optical image (a), AFM image (b) and zoomed-in AFM image (c) of exfoliated pristine $NiPS_3$. d-f) Optical image (d), AFM image (e) and zoomed-in AFM image (f) of exfoliated intercalated $THA-NiPS_3$. Insets in b and e indicate the thickness line profiles of pristine $NiPS_3$ and intercalated $THA-NiPS_3$, respectively.

We investigate whether defects are formed in intercalated $THA-NiPS_3$ by Raman spectroscopy, which is an eminent technique for the characterization of many properties, such as the number of layers, strain, disorder, and defect density, etc., of 2D materials.[48] **Figure 4**a compares the Raman spectra of bulk pristine $NiPS_3$ and bulk intercalated $THA-NiPS_3$. Eight Raman-active phonon modes (three out-of-plane $A_{1g}$ modes and five in-plane $E_g$ modes) are observed in pristine $NiPS_3$, which is consistent with previous results.[16] For intercalated $THA-NiPS_3$, most peaks of pristine $NiPS_3$ are still observed, and a new peak near 206 cm$^{-1}$ that is absent in bulk pristine $NiPS_3$ appears. This peak is ascribed to the resonance-enhanced multi-phonon scattering, and is only observed in ultrathin $NiPS_3$ flakes.[16, 45] The evolution of three out-of-plane modes ( $A_{1g}^1$ , $A_{1g}^2$ , and $A_{1g}^3$ , which are sensitive to interlayer coupling) of $NiPS_3$ after intercalation is consistent with the results of pristine $NiPS_3$ as thickness decreases (Figure S2, Supporting Information).[45] The Raman spectroscopy of intercalated $THA-NiPS_3$ is very similar to that of exfoliated monolayer $NiPS_3$,[16] and no signature related to defects and strain was observed, indicating that the intercalated organic cations effectively reduce interactions between adjacent layers without breaking the in-plane covalent bonds.

We further investigate whether impurity phases are formed in intercalated THA-NiPS₃ by X-ray photoelectron spectroscopy (XPS), which identifies the chemical shift caused by electron state surrounding the atoms of NiPS₃ after intercalation, and the results are shown in Figure 4b. For pristine NiPS₃, the Ni 2p spectrum consists of two main peaks located at binding energies of 854.5 eV and 871.8 eV accompanied with satellite peaks, which correspond to $2p_{3/2}$ and $2p_{1/2}$ levels, respectively. After intercalation, both Ni $2p_{3/2}$ and Ni $2p_{1/2}$ peaks obviously shift towards lower binding energies and the intensities of the satellite peaks are weakened, which is consistent with the results in lithium intercalated NiPS₃ ($Li_xNiPS_3$).[49] Similarly, both P 2p spectrum ($2p_{3/2}$ and $2p_{1/2}$) and S 2p spectrum ($2p_{3/2}$ and $2p_{1/2}$) shift towards lower binding energies after intercalation (Table S1, Supporting Information). No signature indicative of impurity phases was observed, and the shift of XPS peaks towards lower binding energies clearly indicates that the intercalation of THA⁺ cations leads to electron doping.[35]

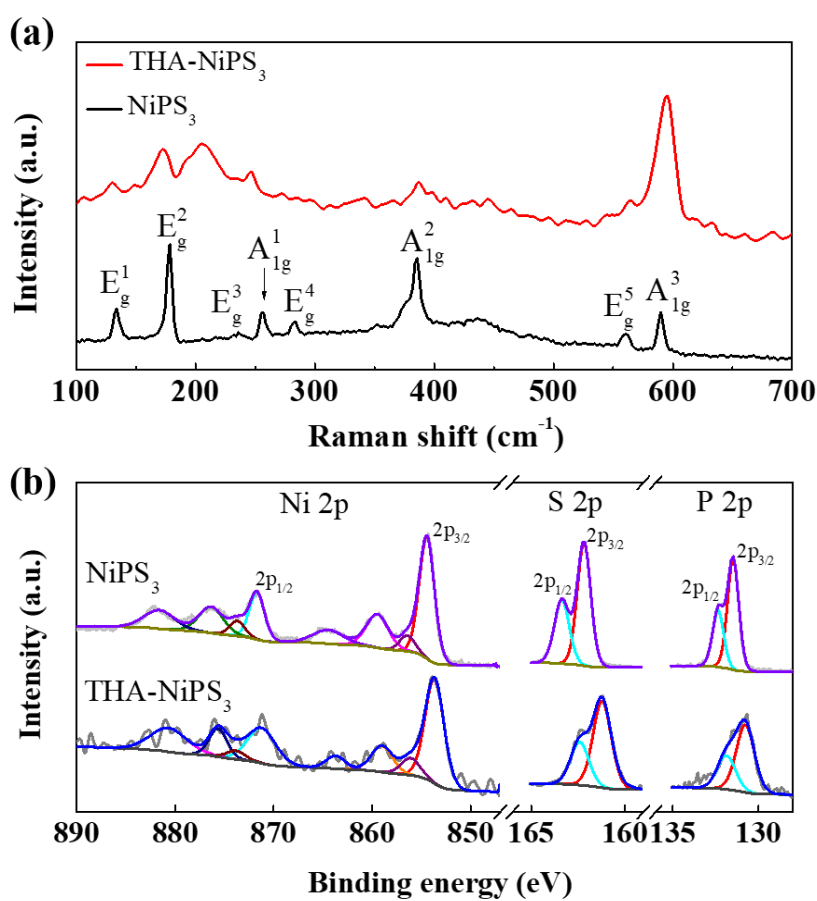

**Figure 4**. Raman (a) and XPS (b) spectra of pristine NiPS₃ and intercalated THA-NiPS₃.

## 2.4 Mechanism for electron doping mediated magnetic orders of NiPS₃

The intercalated THA$^+$ cations decouple the interaction between adjacent NiPS$_3$ layers and lead to electron doping of NiPS$_3$ without introducing obvious strain, defects, and impurity phases. To understand the effects of electron doping on the electronic and magnetic properties of NiPS$_3$, we performed first-principles calculations with Quantum-ESPRESSO.[50] The magnetic ground state of NiPS$_3$ at different doping level is determined by comparing the relative energies of four different magnetic orders (ferromagnetic (FM) order, stripy AFM (sAFM) order, Néel AFM (nAFM) order and zigzag AFM (zAFM) order). As shown in **Figure 5**a, the pristine NiPS$_3$ exhibits zigzag AFM order, where the inner-chain magnetic order is ferromagnetic carrying a magnetic moment of 1.51 $\mu_B$ per Ni atom, and the inter-chain magnetic order is antiferromagnetic, which agrees well with previous studies.[51-52] At a doping concentration of less than 0.5 electrons/cell, the zigzag AFM order still has the lowest energy. At a doping concentration of 0.6 electrons/cell, the Néel AFM order is more stable, while the energy difference between the Néel AFM order and the zigzag AFM order is substantially small.

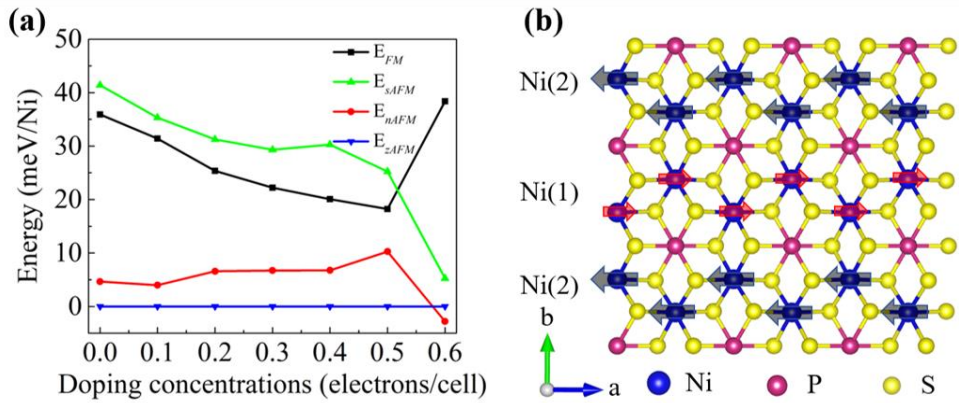

**Figure 5.** a) The relative energies of the four different magnetic orders (FM, sAFM, nAFM, zAFM) as a function of doping concentration. For every doping concentration, the energy of zigzag AFM order is set to zero. Black, green, red and blue lines represent the energies of the FM order, sAFM order, nAFM order and zAFM order, respectively. b) The zigzag AFM order of moderately doped NiPS$_3$ exhibiting FIM characteristics. Arrows indicate the orientation and size of magnetic moments of Ni atoms.

According to the energy differences between the four different magnetic orders, we extract the nearest-neighbor ($J_1$), second-nearest-neighbor ($J_2$), and third-nearest-neighbor ($J_3$) coupling constants (details for calculation in Supporting Information): $J_1$ = -2.50 meV, $J_2$ = -0.26 meV and $J_3$ = 12.90 meV, which are consistent with previous results.[33, 52-54] Among the three magnetic exchange coupling parameters, $J_3$ >>$J_1$ >>$J_2$. $J_3$ is strongly AFM stemming from super-exchange interactions and is the dominant exchange coupling. Therefore, the energies of Néel AFM order and zigzag AFM order are close and are much smaller than the energies of the stripy AFM order and the FM order. In the following, we calculated the band structure and projected density of states (PDOS) of $NiPS_3$ at various doping levels with zigzag AFM order, to focus on the evolution of the electronic structure as a function of doping concentration.

**Figure 6**a shows the calculated band structure of pristine $NiPS_3$, where we could see an isolated narrow conduction band which is dominated by the Ni-d orbitals ($d_{zx}$, $d_{zy}$, and $d_{xy}$) and contributed by S-p orbitals. Without doping, the electronic bands are spin degenerate. Figure 6b shows atom PDOS of $NiPS_3$ with different doping concentrations. With light doping concentration of 0.1 electrons/cell, no spin-splitting occurs on the band. With moderate doping concentration of 0.2-0.5 electrons/cell, we could see obvious splitting on both orbitals and spin states, and the splitting gets larger as doping concentration increases. With further increasing doping concentration to 0.6 electrons/cell, the spin-splitting disappears, while the splitting of orbitals still exists.

To understand the variation of magnetic structures with doping concentrations, we further calculated the orbital-resolved band structure (right panel in Figure 6a) and PDOS of Ni atoms on each ferromagnetic chain with opposite orientation of magnetic moments (Ni(1) site and Ni(2) site) at different doping concentrations (Figure 6c). Interestingly, we could see that at moderate doping concentration of 0.2-0.5 electrons/cell, the splitting causes the doped electrons to occupy Ni atoms at one zigzag ferromagnetic chain. This phenomenon may be understood from Stoner model,[55] which is based on the competition between kinetic energy and exchange energy and is common in material with narrow conduction band.[56-57] From the calculations, we could see that the conduction band minimum (CBM) is dominated by $d_{zx}$, $d_{zy}$, and $d_{xy}$ orbitals of Ni atoms. As doping concentration increases, due to the electron correlation, the $d_{zy}$ orbital is further pushed up and the CBM is then dominated by $d_{zx}$ and $d_{xy}$ orbitals, which are mainly overlapped in the inter-chain direction, and the overlapping is responsible for the doped electrons to be inter-chain itinerant. Further, $NiPS_3$ is a self-doped insulator,[31] the S atoms that separate the Ni chains are not fully occupied. Therefore, after doping electrons, the not fully occupied S-p orbital provides a channel for the electrons to be itinerant between the Ni chains.

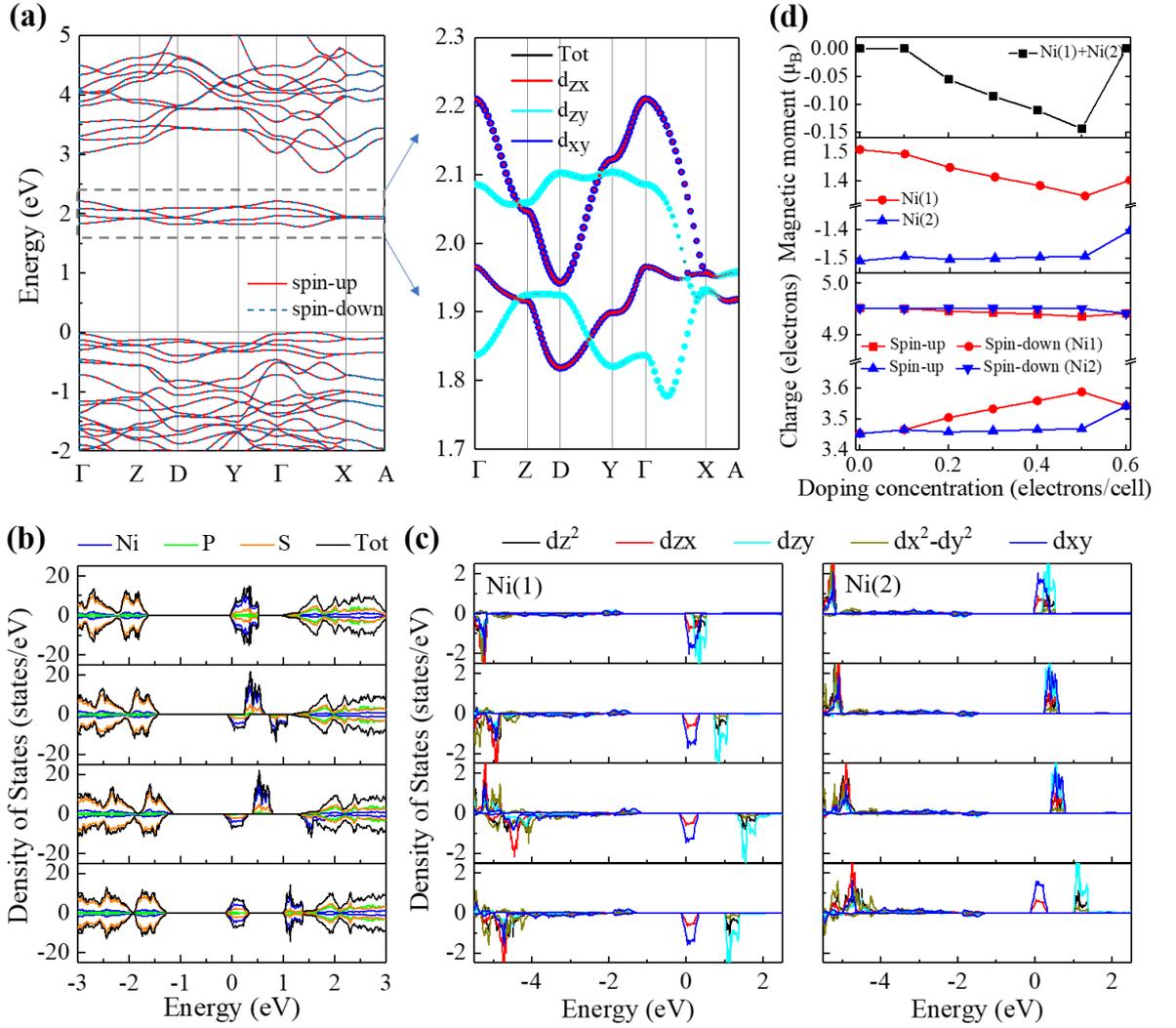

**Figure 6.** Band structures and electronic properties of NiPS$_3$ at different doping concentrations. a) Band structure of spin-up and spin-down configurations of pristine NiPS$_3$ (left) and enlarged region of conduction band minimum (right), the orbital characters of bands are represented by different colors. b) Element-resolved projected density of states (PDOS) of NiPS$_3$ with doping concentrations of 0.1, 0.2, 0.5 and 0.6 electrons/cell from top to bottom. c) Ni's d-orbital resolved PDOS with doping concentrations of 0.1, 0.2, 0.5 and 0.6 electrons/cell from top to bottom of Ni (1) (left) and Ni (2) (right). d) Net magnetic moments (top), magnetic moments of Ni (1) and Ni (2) (middle) and charges of Ni (1) and Ni (2) with spin-up and spin-down configurations (bottom) as a function of doping concentrations. A doping concentration of 0.1 electrons/cell corresponds to an electron density of $0.8 \times 10^{13}$ cm$^{-2}$.

According to Stoner criterion, the spin splitting will occur when $D(E_F) \times I > 1$, where $D(E_F)$ is the total density of states at the Fermi level, and $I$ is the Stoner parameter, which can be estimated from dividing the exchange splitting ($\Delta$) of spin-up and spin-down bands by the corresponding magnetization density ($m$, $m = n^\uparrow - n^\downarrow, n^\uparrow (n^\downarrow)$) is the total number of electrons in spin up (down) band).[55-56] With light doping, even though the electrons are inter-chain itinerant, the small DOS close to the CBM is not able to trigger the Stoner effect. At the doping concentration of 0.2 electrons/cell, $I$ is determined as 1.55 eV and $D(E_F)$ is determined as 10.7 states/eV (Figure S4, Supporting Information), the Stoner criterion is satisfied, giving rise to the inter-chain spin splitting. Thus, the doped electrons only occupy one Ni chain. Figure 6d shows the charges and magnetic moments on Ni(1) and Ni(2) as a function of doping concentration. At the doping concentration of 0.2-0.5 electrons/cell, a net magnetic moment shows up due to the unequal magnetic moments on Ni(1) and Ni(2), which results in the FIM characteristics we found in experiment in intercalated THA-NiPS$_3$. Figure 5b shows a schematic of the magnetic order of the moderately doped NiPS$_3$ exhibiting FIM characteristics, where the antiferromagnetically coupled two zigzag ferromagnetic chains carry unequal magnetic moments.

However, when the doping concentration further increases to 0.6 electrons/cell, net magnetic moment disappears (Figure 6d) and the system prefers the Néel AFM order (Figure 5a). The ground magnetic order is determined by the competition between Stoner exchange dominated FM order and super-exchange dominated AFM order, and the super-exchange starts to dominate the magnetic order at the doping concentration of 0.6 electrons/cell. Quantitatively understanding the magnetic order transition at high doping concentration needs to be exploited further in the future.

We further quantitatively compare the experimental results with the calculated theoretical values. The observed FIM signal of THA-NiPS$_3$ is weak, and the corresponding average net magnetic moment per cell is approximately 0.07 $\mu_B$, which lies in the range of theoretically expected values. Due to the highly insulating property of NiPS$_3$ (conductivity ~$10^{-7}$ S cm$^{-1}$ at room temperature[45]) and degradation of intercalated NiPS$_3$ during device fabrication process, quantitative determining the carrier density in intercalated NiPS$_3$ is challenging. Previous studies have shown that intercalated THA$^+$ cations cause a doping concentration of 0.02 electrons per MoS$_2$ formula unit[37] (corresponding to a carrier density of 2.3×10$^{13}$ cm$^{-2}$), assuming that intercalated THA$^+$ cations cause a similar doping concentration in NiPS$_3$, which lies in the appropriate doping concentration that leads to the net magnetic moment.

Depending on the size and arrangement of organic cations, both the doping concentration and the interlayer distance can be tuned by intercalating various organic cations into the vdW gaps of NiPS$_3$. For example, cations such as tetrabutyl ammonium (TBA$^+$), tetrapropyl ammonium (TPA$^+$), and cetyltrimethyl ammonium (CTA$^+$) cause a substantially large doping concentration (>$10^{14}$ cm$^{-2}$) in intercalated 2D materials.[28, 37, 39] The conductivity of NiPS$_3$ increases with increasing doping concentration in lithium intercalated NiPS$_3$.[58] We measured conductivities of NiPS$_3$ intercalated with various organic cations (Figure S6, Supporting Information). After intercalation, the conductivity of NiPS$_3$ increases, and the conductivities of TBA-NiPS$_3$, TPA-NiPS$_3$ and CTA-NiPS$_3$ are significantly greater than that of THA-NiPS$_3$, confirming that intercalated TBA$^+$, TPA$^+$, and CTA$^+$ cations cause a greater doping concentration than intercalated THA$^+$ cations. Similar to THA-NiPS$_3$, the interlayer distance increases to 11.46 Å, 12.14 Å and 14.82 Å for TBA-NiPS$_3$, TPA-NiPS$_3$ and CTA-NiPS$_3$ (as shown in **Figure 7**a), respectively, and the Raman spectra of TBA-NiPS$_3$, TPA-NiPS$_3$ and CTA-NiPS$_3$ also exhibit characteristics of monolayer NiPS$_3$ sheet (Figure S6a, Supporting Information). The interlayer distance of every of the above organic cations intercalated NiPS$_3$ is almost 2 times of the interlayer distance of the pristine NiPS$_3$. As the interlayer distance of NiPS$_3$ increases, the coupling interaction between adjacent layers decreases rapidly and even reaches zero (Figure S7, Supporting Information), indicating that the interlayer magnetic coupling in intercalated NiPS$_3$ is negligible. As predicted by our theoretical model that the net magnetic moment disappears at high doping concentration, we do not observe any FIM characteristics in the heavily doped NiPS$_3$ samples, except a linear dependence of *M vs H* (as shown in Figure 7b). These results further rule out the possibility of tuning magnetic order transition from AFM to FIM in intercalated NiPS$_3$ by interlayer distance, and confirm the role of carrier doping in tuning the magnetic properties of NiPS$_3$.

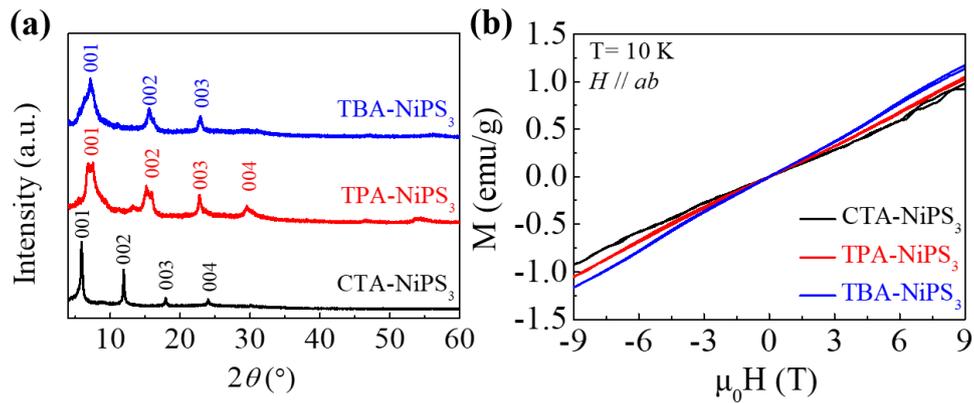

**Figure 7.** XRD patterns (a) and field dependence of magnetization (*M-H*, b) measured under magnetic field *H // ab* at 10 K of intercalated TBA-NiPS$_3$, TPA-NiPS$_3$, and CTA-NiPS$_3$.

## 3. Conclusion

We successfully realized the AFM-FIM-AFM transition in $NiPS_3$ with carrier doping by intercalating organic cations into vdW gaps of $NiPS_3$. The intercalated organic cations decouple interactions between adjacent layers without introducing defects and impurity phases, and result in electron doping, which significantly alters the electronic and magnetic properties of $NiPS_3$. At an appropriate doping concentration (which is also achievable by electrostatic gating method), the AFM order in pristine $NiPS_3$ is switched to the FIM order with $T_c = 100$ K in THA-$NiPS_3$. At heavy doping concentration (TBA-$NiPS_3$, TPA-$NiPS_3$ and CTA-$NiPS_3$), the AFM order remains again. Such carrier doping tuned magnetic transition arises from the competition between Stoner exchange and super-exchange as the variation of the doping concentration. Our work provides a viable tool to modulate the magnetic properties of vdW magnets by electrical method, and opens a way for investigating strong correlation between electronic structure and magnetic properties of the vdW magnets and designing novel spintronic devices.

## 4. Experimental Section

Sample preparation: $NiPS_3$ crystals were grown by vapor transport method. The mixture of stoichiometric high-purity Ni, P, S (Ni/P/S = 1:1:3) and iodine (10 mg cm$^{-3}$) as a transport agent were sealed into an evacuated quartz ampule and kept in a two-zone furnace (650 °C - 600 °C) for 1 week. The large single crystals will be harvested in the lower temperature end. Thin flakes of both pristine $NiPS_3$ and intercalated $NiPS_3$ were prepared on silicon substrate with a layer of 285 nm $SiO_2$ by mechanical exfoliation from bulk crystal using adhesive tape.

Electrochemical intercalation: The electrochemical intercalation of $NiPS_3$ was carried out in a two-electrode system. The fresh $NiPS_3$ crystal was fixed on an electrode holder as the negative electrode (cathode), a piece of Pt was used as the positive electrode (anode), and tetraheptyl ammonium bromide (THA$^+$Br$^-$, 0.1 g, Macklin, 98%) dissolved in acetonitrile (20 ml, Macklin, 99%) was used as electrolyte. During the intercalation process, the voltage was slowly swept from 0 V to ~4 V at 50 °C. The electrochemical reaction consists of two half-reactions:

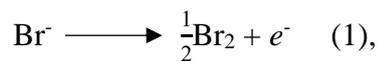
$$Br^- \longrightarrow \tfrac{1}{2}Br_2 + e^- \quad (1),$$

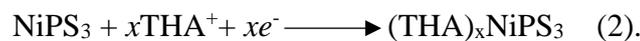
$$NiPS_3 + xTHA^+ + xe^- \longrightarrow (THA)_xNiPS_3 \quad (2).$$

Two bromide ions lose electrons to form $Br_2$ at the anode, THA$^+$ cations are inserted into the vdW gaps of $NiPS_3$ (cathode), and $NiPS_3$ receives electrons from the external circuit. Therefore, organic cations intercalation causes electron doping of the sample. Intercalation of other organic

WILEY-VCH

cations, such as tetrabutyl ammonium (TBA$^+$), tetrapropyl ammonium (TPA$^+$), and cetyltrimethyl ammonium (CTA$^+$), share the same procedure as the THA$^+$.

Characterization: The X-ray diffraction (XRD) patterns were collected by SmartLab® high-resolution X-ray diffractometer (Rigaku, Japan) using Cu $K_\alpha$ radiation, $\lambda$ = 1.5418 Å. The X-ray photoelectron spectroscopy (XPS) spectra were obtained in Nano-X by PHI-5000 VersaprobeII (Ulvac-Phi, Japan) using Al $K_\alpha$ X-ray ($hv$ = 1486.6 eV). The binding energies (BE) were calibrated with respect to the C-C 1s bond (BE = 284.8 eV). Raman spectra were collected by inVia™ confocal Raman microscope (Renishaw) using an excitation wavelength of 532 nm. The magnetic properties were measured by a vibrating sample magnetometer (VSM) of a commercial physical property measurement system (Dynacool-9, Quantum Design).

First-principles density functional theory (DFT) calculations: The Vanderbilt ultra-soft pseudopotential within exchange-correlation potential described by GGA+U, where U = 6 eV is applied on the Ni site.[59] We use the correction of DFT-D to consider the van der Waals interactions between layers.[60] For bulk computations, we construct a supercell of 20 atoms, and 8×5×7 k points mesh generated by Monkhorst-Pack scheme is used. The cutoff kinetic energy of 50 Ry (500 Ry) for wave function (charge density) is employed for all computations. Convergence with respect to the plane-wave cutoff energy and k-point sampling has been carefully checked. The structure optimizations are performed until the force on each atom is less than $10^{-4}$ Ry/Bohr and the convergence threshold for self-consistency is $10^{-8}$ Ry. The carrier doping is tuned by changing the total number of electrons adding in the cell, with a compensating jellium background of opposite charge to maintain charge neutrality.

**Supporting Information**

Supporting Information is available from the Wiley Online Library or from the author.

**Acknowledgements**

M.M. X.Z. and S.W. contributed equally to this work. This work was supported by the National Natural Science Foundation of China (Grant Nos. 92065206, 11974211), the Natural Science Foundation of Shandong Province (Grant No. ZR2020MA071), and the Qilu Young Scholar Program of Shandong University.

**Conflict of Interest**

The authors declare no competing interests.